\documentclass[prl,twocolumn,showpacs,groupaddress,floatfix,superscriptaddress]{revtex4-1}
\usepackage{graphicx,graphics}
\usepackage{dcolumn}
\usepackage{amsmath,amssymb,amsfonts}
\usepackage{latexsym,verbatim}
\usepackage{bm}
\usepackage{color}
\usepackage[breaklinks=true,colorlinks,citecolor=blue,linkcolor=blue,urlcolor=blue]{hyperref}

\usepackage{fancybox}
\usepackage{tabularx}

\begin{document}

\author{Alessandro Principi}
\affiliation{Radboud University, institute for Molecules and Materials, NL-6525 AJ Nijmegen, The Netherlands}

\author{Mikhail I. Katsnelson}
\affiliation{Radboud University, institute for Molecules and Materials, NL-6525 AJ Nijmegen, The Netherlands}

\author{Alex Levchenko}
\affiliation{Department of Physics, University of Wisconsin-Madison, Madison, Wisconsin 53706, USA}

\title{Chiral second-sound collective mode at the edge of 2D systems with nontrivial Berry curvature}

\begin{abstract}
We study the thermal transport in two-dimensional systems with a nontrivial Berry curvature texture. The physical realizations are many: for a sake of definiteness we consider undoped graphene gapped by the presence of an aligned hexagonal-Boron-Nitride substrate. The same phenomenology applies, i.e., to surface states of 3D topological insulators in the presence of a uniform magnetization. We find that chiral valley-polarized second-sound collective modes propagate along the edges of the system. The localization length of the edge modes has topological origin stemming from the anomalous velocity term in the quasiparticle current. At low temperature, the single-particle contribution to the transverse thermal conductance is exponentially suppressed and only second-sound modes carry heat along the boundary. A sharp change in the behavior of the thermal Hall conductance, extracted from nonlocal measurements of the temperature along the edge, marks the onset of ballistic heat transport due to second-sound edge modes.
\end{abstract}

\date{\today}


\maketitle

{\it Introduction}---Managing heat production and transfer is one of the major challenges of the present time~\cite{Gammaitoni_nanotech_2015}. The constant miniaturization of the electronic circuitry heavily relies on the reduction of the heat produced by a single element or, alternatively, on efficient dissipation mechanisms. The quest for novel materials with either of these properties has become more vibrant in recent years, fueled by the discovery of two-dimensional (2D) materials~\cite{Novoselov_science_2004,Novoselov_pnas_2005}. Among these, graphene stands out for its record-high thermal conductivities~\cite{Balandin_nanolett_2008,Ghosh_apl_2008,Lee_prb_2011,Balandin_nature_mater_2011}. 

Although materials are new, the theory describing heat conduction in solids is well established~\cite{Ziman_book,Ashcroft_Mermin,Landau_Lifshitz_10}. Both electrons and phonons contribute to the transfer of heat from hot to cold regions~\cite{Ziman_book,Ashcroft_Mermin,Landau_Lifshitz_10}.
While in insulators and semiconductors phonons dominate the heat transport, in metals and semimetals the contribution of electrons can be comparable, or even dominant over that of lattice vibrations~\cite{Ziman_book}. In this paper we focus on the transfer of heat due to electrons in the hydrodynamic regime~\cite{Landau_Lifshitz_6,Giuliani_Vignale,Torre_prb_2015,Principi_prb_2016,Bandurin_science_2016,Crossno_science_2016}.  

Usually, electron-impurity scattering occurs at a much higher rate than electron-electron collisions~\cite{Giuliani_Vignale}. In this case, a ``diffusive'' regime is established.
The evolution of the temperature is well described by the Fourier law of heat conduction~\cite{Landau_Lifshitz_6,Landau_Lifshitz_10}. Two transport quantities control the density and heat diffusion equations, namely the charge and thermal conductivities. 
When the system is in the Fermi-liquid regime, they are proportional to each other. This property, the so-called Wiedeman-Franz law~\cite{Ziman_book}, reflects the fact (i) that the same kind of quasiparticle carries both charge and heat, and (ii) that the scattering mechanism affects in the same way the two transport channels~\cite{Ashcroft_Mermin,Landau_Lifshitz_10}. 
However, even weak electron-electron interactions modify this result~\cite{Castellani_prb_1986,Castellani_prl_1987,Arfi_journ_low_t_phys_1992,Schwab_ann_of_phys_2003,Raimondi_prb_2004,Catelani_JETP_2005,Schwiete_JETP_2016}. 

When a system is strongly interacting and ultra clean  
the whole picture of transport changes dramatically:
the electron liquid is driven into the hydrodynamic regime~\cite{Giuliani_Vignale,Andreev_prl_2011,Muller_prl_2009,Principi_prb_2015,Narozhny_prb_2015,Briskot_prb_2015}. Electrons are described by a local-equilibrium distribution function, {\it i.e.} by a Fermi distribution whose chemical potential and temperature are space- and time-dependent and whose momentum is shifted by the local value of the drift velocity of the fluid~\cite{Landau_Lifshitz_10}. These three quantities are determined from the knowledge of the local values of the densities of the conserved quantities, {\it i.e.} the particle number, energy and momentum, respectively.
Their equations of motion, {\it i.e.} the continuity, Navier-Stokes and heat-conduction equations, are controlled by a handful of parameters that can be calculated from the microscopic model~\cite{Landau_Lifshitz_6,Principi_prb_2015,Principi_prl_2015,Principi_prb_2016}.

Notably, the hydrodynamic equations admit solutions in which the heat is transferred ballistically by means of entropy waves, {\it i.e.} the so-called second sound~\cite{Landau_Lifshitz_6,Andreev_prl_2011}. The cleanest observation of second sound occurred in ${}^4{\rm He}$ below the $\lambda$-point (corresponding to a temperature of $\sim 2.18~{\rm K}$~\cite{Ackerman_prl_1966}). Few other systems have been reported to support ballistic heat transfer~\cite{Ackerman_prl_1969,Narayanamurti_prl_1972,Jackson_prl_1970}. Recently, undoped graphene was proposed as a possible material in which this phenomenon could be observed~\cite{Phan_arxiv_2013}. 

When the inversion symmetry of the hexagonal lattice is broken and a gap is opened at the Dirac point, the system exhibits hotspots of Berry curvature at the two inequivalent valleys (${\bm K}$ and ${\bm K}'$) of the Brillouin zone~\cite{Shapere_Wilczek,Gorbachev_science_2014}. The latter acts as a gauge field in momentum space, and leads to observable physical effects~\cite{Culcer_prl_2004,Xiao_prl_2006,Xiao_prl_2007,Xiao_prb_2008,Qin_prl_2011,Gorbachev_science_2014,Pellegrino_prb_2015}. When an external force due, e.g., to an electric field or thermal gradient sets electrons in motion, they ``skew'' in the orthogonal direction under the effect of the Berry curvature and transverse (electric or thermal) currents appear~\cite{Gorbachev_science_2014}. Since the time-reversal symmetry is unbroken, electrons in different valleys experience Berry curvatures with opposite signs~\cite{Xiao_rmp_2010}. Two counter-propagating transverse currents appear, each of them due to the electrons of one valley. The net current therefore vanishes, but the ``valley'' current stays finite~\cite{Xiao_prl_2007,Gorbachev_science_2014}. 

The Berry curvature (${\bm \Omega}_{\bm k}$) has also an important impact on the collective modes of the electron liquid~\cite{Haldane_prl_2004,Son_prl_2012,Lundgren_prl_2015,Song_pnas_2016}, 
{\it i.e.} the self-sustaining solutions of the coupled Navier-Stokes and Maxwell's equations (the latter are needed only to determine the charged modes~\cite{Giuliani_Vignale}). Notably, in two dimensions and in the absence of a magnetic field, the Berry curvature does not appear explicitly in the linearized equations describing the motion of the fluid in the bulk. However, it crucially enters into the boundary conditions, and it has been shown to stabilize charged collective modes localized at the edges of the sample (edge plasmons~\cite{Song_pnas_2016}). When an electric field is applied, the quasiparticle velocity acquires an anomalous component, orthogonal to both ${\bm \Omega}_{\bm k}$ and the applied force, which adds up to the Bloch band velocity~\cite{Xiao_rmp_2010}. Since at the boundaries the {\it total} velocity perpendicular to the edges has to vanish, the presence of the anomalous term leads to non-trivial solutions of Navier-Stokes equations~\cite{Song_pnas_2016}. These solutions would not exist in the absence of the anomalous velocity (or of an external magnetic field~\cite{Volkov_jetp_1988,Fetter_prb_1985,MacDonald_prl_1985,Fetter_prb_1986}). 

\begin{figure}[t]
\begin{center}
\begin{tabularx}{\columnwidth}{X}
\noindent\parbox[c]{\hsize}{\includegraphics[width=0.99\columnwidth]{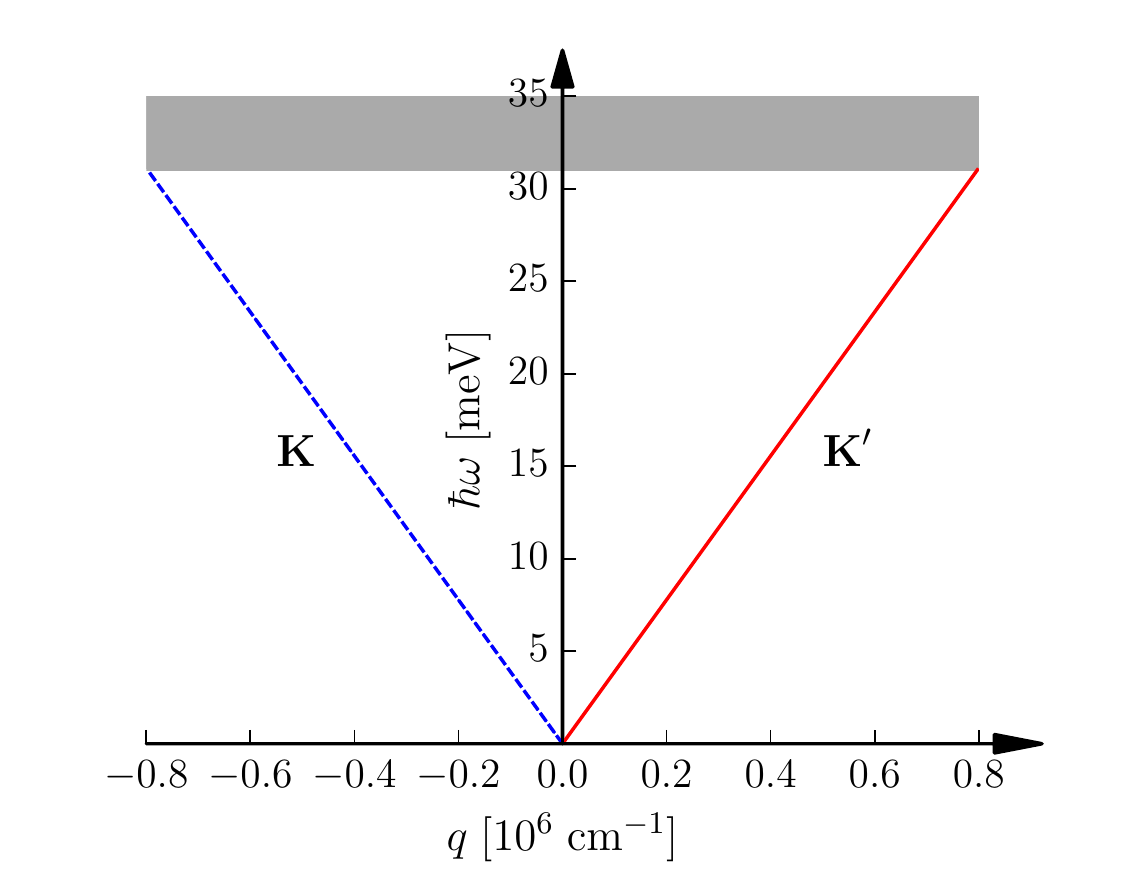}}
\end{tabularx}
\end{center}
\caption{
The edge second sound dispersion for a half-gap $\Delta/k_{\rm B} = 180~{\rm K}$ and a temperature $T=90~{\rm K}$ [$k_{\rm B} T/(2\Delta) = 1/4$]. The shaded region correspond to the particle-hole continuum ($\hbar \omega > 2 \Delta$). Note that the edge second-sound modes are completely valley polarized.
\label{fig:one}}
\end{figure}

In this paper we study the problem of second-sound in graphene, when inversion symmetry is broken by e.g. the presence of an aligned hexagonal Boron Nitride (hBN) substrate~\cite{Hunt_science_2013,Dean_nature_2013,Gorbachev_science_2014}. We focus on the physics around the Dirac points, neglecting the effect of mini-Dirac cones stemming from the Moir\'e superlattice~\cite{Ponomarenko_nature_2013,Yankowitz_nature_phys_2012}. We show that, when the system is undoped and the temperature is smaller than the band gap, it cannot support edge plasmons but only charge-neutral second-sound collective modes localized at the boundary [see Fig.~\ref{fig:one}]. Notably, since the electrons of each valley experience opposite Berry curvatures, {\it two} counter-propagating and completely valley-polarized second-sound edge modes are found. We discuss the experimental conditions and setups under which these can actually be observed.

%%%%%%%%%%%%%%%%%%%
%%%%%%%%%%%%%%%%%%%

{\it Model}---We consider electrons in gapped graphene in the presence of thermal fluctuations. The Hamiltonian reads (hereafter $\hbar = 1$)~\cite{Castro_Neto_rmp_2009,Goerbig_rmp_2011,Das_Sarma_rmp_2011,Kotov_rmp_2012,Grigorenko_nat_phot_2012}
\begin{eqnarray}
{\cal H} = \sum_{{\bm k},\alpha,\beta} c^\dagger_{{\bm k},\alpha} {\cal E}_{\bm k} \cdot {\bm \sigma}_{\alpha\beta} c_{{\bm k},\beta}
~,
\end{eqnarray}
where $c^\dagger_{{\bm k},\alpha}$ ($c_{{\bm k},\alpha}$) creates (destroys) an electron with momentum ${\bm k}$  
in sublattice $\alpha=A,B$, ${\cal E}_{\bm k} = (v_{\rm F} k_x, v_{\rm F} k_y, \Delta)$, $v_{\rm F}$ is the Fermi velocity, and $\Delta$ is (half) the band gap. For graphene on hBN the band gap  
is predicted to vary between $10~{\rm meV}$ and $300~{\rm meV}$, depending e.g. on the misalignment angle between the two 
structures~\cite{Woods_nat_phys_2014,Titov_prl_2014,Bokdam_prb_2014,Slotman_prl_2015}. For our estimates we use the fairly small value $\Delta/k_{\rm B} \simeq 180~{\rm K}$, which was reported in recent experiments performed on a particular sample of graphene on hBN~\cite{Woods_nat_phys_2014,Gorbachev_science_2014}. Naturally, if the value of $\Delta$ is larger (as it is theoretically predicted for a perfect alignment), we can expect our findings to be relevant also for room-temperature experiments. The band energy is $\varepsilon_{{\bm k},\lambda} = \lambda\sqrt{(v_{\rm F} k)^2 + \Delta^2}$ ($\lambda = \pm$ label the conduction and valence bands), while the Berry curvature in valley ${\bm K}$ is~\cite{Xiao_rmp_2010}
\begin{eqnarray}
{\bm \Omega}_{{\bm k},\lambda} = - \lambda \frac{v_{\rm F}^2 \Delta}{2[(v_{\rm F} k)^2+\Delta^2]^{3/2}} {\hat {\bm z}}
~.
\end{eqnarray}
Because of the time-reversal symmetry, the Berry curvature of the valley ${\bm K}'$ has the opposite sign ({\it i.e.} its band gap is $-\Delta$). In the presence of an electric field ${\bm E}({\bm r},t)$, the quasiparticle velocity becomes ${\bm v}_{{\bm k},\lambda} = {\bm \nabla}_{{\bm k}} \varepsilon_{{\bm k},\lambda} + {\bm E}\times {\bm \Omega}_{{\bm k},\lambda}$. The second term is the so-called anomalous velocity. Hydrodynamic equations are derived by assuming that the system is in local quasi-equilibrium with the distribution function~\cite{Haldane_prl_2004,Son_prl_2012,Lundgren_prl_2015,Song_pnas_2016}
\begin{eqnarray} \label{eq:f_k_hydro}
f_{{\bm k},\lambda} = \big\{1 + \exp[(\varepsilon_{{\bm k},\lambda} - {\bm u}\cdot{\bm k} - {\tilde \mu})/(k_{\rm B} {\tilde T})]\big\}^{-1}
~,
\end{eqnarray}
where ${\tilde T} = T + \delta T({\bm r},t)$, ${\tilde \mu} = \mu + \delta\mu({\bm r},t)$ and ${\bm u} = {\bm u}({\bm r},t)$ are, respectively, the local temperature, chemical potential and fluid velocity, while $k_{\rm B}$ is the Boltzmann constant. We define $T$ and $\mu$ as the (uniform) equilibrium temperature and chemical potential. 
%Their variations are hereafter called $\delta T$ and $\delta \mu$, and throughout the paper 
We will assume that $|\delta T| \ll T$ and $|\delta\mu| \ll {\rm max}(|\mu|, \Delta)$. The Navier-Stokes equations, obtained by integrating the semiclassical Boltzmann equation over its momenta ($1, {\bm k}, \varepsilon_{{\bm k},\lambda}-\mu$), read (we suppress space and time indices)~\cite{Landau_Lifshitz_6,Principi_prb_2016,Narozhny_prb_2015,Briskot_prb_2015}
\begin{eqnarray} \label{eq:NS_equations}
\left\{
\begin{array}{l}
\partial_t n + {\bm \nabla}\cdot {\bm j}_n = 0
\vspace{0.2cm}\\
\partial_t {\bm p} + {\bm \nabla}_i \varsigma_{ij} + e\sigma{\bm E} = 0
\vspace{0.2cm}\\
\partial_t s + {\bm \nabla}\cdot {\bm j}_q = 0
\end{array}
\right.
~.
\end{eqnarray}
Here $n({\bm r},t)$, $s({\bm r},t)$ and ${\bm p}({\bm r},t)$ are, respectively, the number, entropy and momentum densities, while ${\bm j}_n({\bm r},t)$, ${\bm j}_q({\bm r},t)$ and $\varsigma_{ij}({\bm r},t)$ are the corresponding currents. The stress tensor $\varsigma_{ij}$ contains also the kinetic contribution to the equation of motion of the momentum density. Finally, $\sigma$ is the momentum conductance. 
%The role of $\delta T$, $\delta \mu$ and ${\bm u}$ is to introduce at a formal level local fluctuations of the entropy, number, and momentum density, respectively. Once these

Collective modes are found by solving the system of equations~(\ref{eq:NS_equations}) together with the appropriate boundary conditions. We require the component of the particle and heat currents~\cite{Xiao_prl_2006,Qin_prl_2011},
\begin{subequations} \label{eq:boundary_cond}
\begin{eqnarray} 
\label{eq:boundary_cond_jn}
{\bm j}_n &\equiv& \sum_{{\bm k},\lambda} {\bm v}_{{\bm k},\lambda} f_{{\bm k},\lambda} + {\bm \nabla}\times \sum_{{\bm k},\lambda} \xi_{{\bm k},\lambda} {\bm \Omega}_{{\bm k},\lambda} f_{{\bm k},\lambda}
%{\bm j}_n &\equiv& \sum_{{\bm k},\lambda} {\dot {\bm r}}_{{\bm k},\lambda} f_{{\bm k},\lambda} + {\bm \nabla}\times \sum_{{\bm k},\lambda} \xi_{{\bm k},\lambda} {\bm \Omega}_{{\bm k},\lambda} f_{{\bm k},\lambda}
~,
\\
\label{eq:boundary_cond_jq}
{\bm j}_q &\equiv& \sum_{{\bm k},\lambda} \xi_{{\bm k},\lambda} {\bm v}_{{\bm k},\lambda} f_{{\bm k},\lambda}  + {\bm \nabla}\times \sum_{{\bm k},\lambda}  \xi_{{\bm k},\lambda}^2  {\bm \Omega}_{{\bm k},\lambda} f_{{\bm k},\lambda}
%{\bm j}_q &\equiv& \sum_{{\bm k},\lambda} \xi_{{\bm k},\lambda} {\dot {\bm r}}_{{\bm k},\lambda} f_{{\bm k},\lambda}  + {\bm \nabla}\times \sum_{{\bm k},\lambda}  \xi_{{\bm k},\lambda}^2  {\bm \Omega}_{{\bm k},\lambda} f_{{\bm k},\lambda}
~,
\end{eqnarray}
\end{subequations}
perpendicular to the edge to vanish at the edge itself. 
%{\bf Since the momentum is proportional to the energy current ${\bm j}_e = {\bm j}_q + \mu {\bm j}_n$, Eqs.~(\ref{eq:boundary_cond_jn})-(\ref{eq:boundary_cond_jq}) together imply that also the momentum perpendicular to the edge vanishes.}
Here $\xi_{{\bm k},\lambda} = \varepsilon_{{\bm k},\lambda} - \mu$. Note that the last term in Eq.~(\ref{eq:boundary_cond_jq}) can be rewritten as ${\bm j}_{q}^{({\rm tr})} = k_{xy}^{\rm sp} {\hat {\bm z}} \times {\bm \nabla} T$, where $k_{xy}^{\rm sp}$ is the single-particle transverse thermal conductance~\cite{Qin_prl_2011}.

%%%%%%%%%%%%%%%%%%%
\begin{figure}[t]
\begin{center}
\begin{tabularx}{\columnwidth}{X}
\noindent\parbox[c]{\hsize}{\includegraphics[width=0.99\columnwidth]{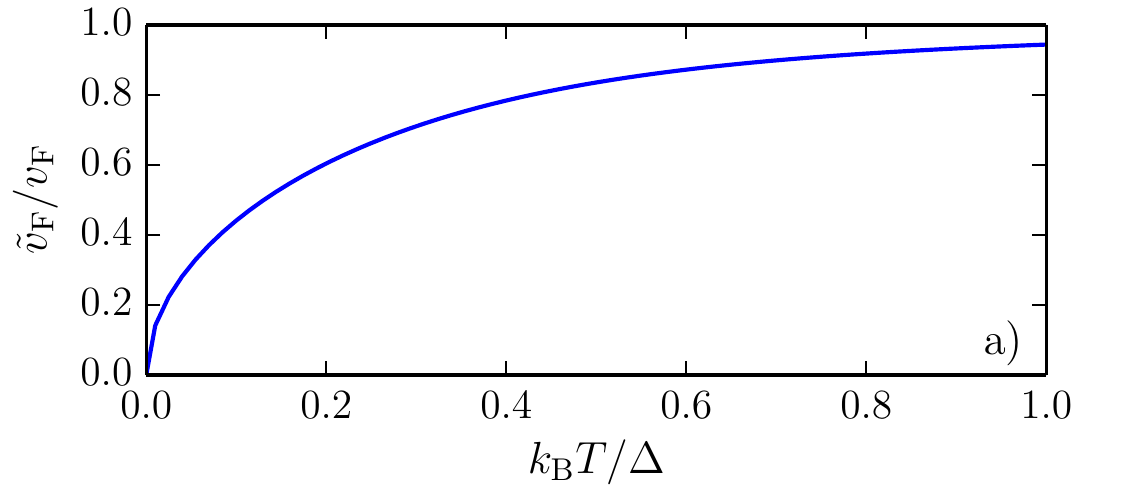}}
\\
\noindent\parbox[c]{\hsize}{\includegraphics[width=0.99\columnwidth]{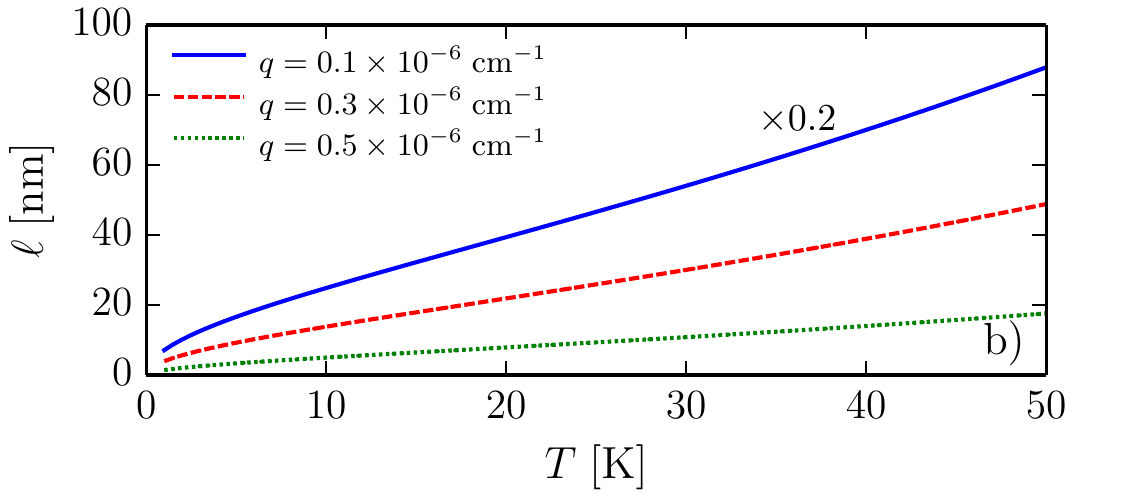}}
\end{tabularx}
\end{center}
\caption{
Panel (a) the velocity ${\tilde v}_{\rm F}$, in units of the Fermi velocity $v_{\rm F}$, as a function of temperature (in units of the half-gap $\Delta$).
Panel (b) The localization length in the ${\hat {\bm x}}$ direction in ${\rm nm}$, plotted as a function of the temperature for three values of the wavevector $q$. Note that the solid curve, corresponding to $q=0.1\times 10^6~{\rm cm}^{-1}$ has been rescaled by multiplying it with a factor $0.2$.
\label{fig:two}}
\end{figure}
%%%%%%%%%%%%%%%%%%%

{\it Second sound in the undoped limit}---In the undoped regime, our description breaks down if the semiclassical electron wave packets are shared between the bands~\cite{Xiao_rmp_2010}. A more complicated non-Abelian description would be needed~\cite{Xiao_rmp_2010}. To keep the presentation as simple as possible, we assume that the gap $2\Delta$ is the largest energy scale, {\it i.e.} we restrict to the case $k_{\rm B} T < 2 \Delta$ (we anyway focus on the case $\{v_{\rm F} q, \omega\} \ll 2\Delta$). In this temperature range interband transitions are exponentially suppressed, and it still makes sense to consider the electronic wave packets as composed by electrons of only one band.

When the chemical potential is exactly in the middle of the gap and in the absence of external electric fields, the continuity equation becomes trivial, since $n= {\bm j}_n = 0$. Moreover, the density remains constant and no self-induced electric field appears. Therefore, we set ${\bm E}=0$ and we consider the last two Navier-Stokes equations, which read (we set $k_{\rm B} = 1$ for convenience)
\begin{eqnarray} \label{eq:NS_undoped}
\left\{
\begin{array}{l}
{\displaystyle \partial_t { {\bm p}} + \frac{T}{2} A(\Delta/T) {\bm \nabla} { s} + \nu {\bm \nabla}^2 {\bm p} = 0 }
\vspace{0.2cm}\\
{\displaystyle \partial_t { s} + \frac{v_{\rm F}^2}{T} {\bm \nabla}\cdot{ {\bm p}} = 0 }
\end{array}
\right.
~,
\end{eqnarray}
where $\nu$ is the kinematic viscosity, $A(x) = 1 - x^2 f_{1}(x)/f_{3}(x)$, and
\begin{eqnarray}
f_{n}(x) = \frac{1}{2}\int_{|x|}^{\infty} dy \frac{y^n}{\cosh^2(y/2)}
~.
\end{eqnarray}
These equations agree with those of Ref.~\cite{Phan_arxiv_2013} in the limit $\Delta \to 0$, and lead to bulk modes with acoustic dispersion $\omega = {\tilde v}_{\rm F} q/\sqrt{2}$, where ${\tilde v}_{\rm F} = v_{\rm F} A^{1/2}(\Delta/T)$
Note that $A(x)$ is always positive and vanishes in the limit $x\to \infty$ [see Fig.~(\ref{fig:two}a)] as $A(x)\to 2/x$.

Let us now consider collective modes localized at the edge. We neglect for the time being the kinematic viscosity $\nu$, which we will reintroduce at the end of the calculation. Indeed, although it is possible to retain it at all steps, the intermediate equations become quite cumbersome. We look for solutions of Eq.~(\ref{eq:NS_undoped}) of the form $s({\bm r},t) = s_0 \exp(x/\ell + i q y -i\omega t)$ and ${\bm p}({\bm r},t) = p_0 \exp(x/\ell + i q y -i\omega t)$, assuming that the system occupies the half space $x<0$. 
We get
\begin{eqnarray} \label{eq:NS_equation_0}
\left\{
\begin{array}{l}
{\displaystyle \omega^2 - {\tilde v}_{\rm F}^2(q^2 - \ell^{-2})/2 = 0 }
\vspace{0.2cm}\\
{\displaystyle -i\omega p_{0,x} + \frac{T}{2} A(\Delta/T) \ell^{-1} s_0 = 0 }
\end{array}
\right.
~.
\end{eqnarray}
Note that it is not $p_{0,x}$, but the sum of $p_{0,x}$ and an anomalous term analogous to those of Eqs.~(\ref{eq:boundary_cond_jn})-(\ref{eq:boundary_cond_jq}) that vanishes at the boundary. This condition is equivalent to require that the {\it physical} energy current ${\bm j}_e = {\bm j}_q + \mu {\bm j}_n$ (which coincides, apart from numerical factors, with the physical momentum density) across the edge vanishes.
%to which we have to add
Eq.~(\ref{eq:NS_equation_0}) must be solved together with
the boundary conditions $j_{n,x}(x=0) = 0$ and $j_{q,x}(x=0) = 0$. The currents are given in Eq.~(\ref{eq:boundary_cond}). It turns out that~(\ref{eq:boundary_cond_jn}) is always zero. Indeed, both the band and anomalous velocities vanish when integrated over both bands. Therefore, we are left to study Eq.~(\ref{eq:boundary_cond_jq}). After some transformations, combining it with the second line of Eq.~(\ref{eq:NS_equation_0}), we get
\begin{equation} \label{eq:bound_cond}
j_{q,x}(x=0) 
=
-i \frac{s_0 v_{\rm F}^2}{2}  \left[ \frac{T}{\omega} A(\Delta/T) \ell^{-1} + q B(\Delta/T) \right]
~,
\end{equation}
where $B(x) = x f_{1}(x)/f_{3}(x)$. In the limit $x\to \infty$, $B(x) \to 1/x$. We remark that the first (second) term inside the square brackets of Eq.~(\ref{eq:bound_cond}) corresponds to the normal (anomalous) component of the heat current. The last term on the right-hand side of this equation was obtained from the anomalous part of Eq.~(\ref{eq:boundary_cond_jq}) [$j_{q,x}^{\rm tr} = -k_{xy}^{\rm sp} \partial_y T$, with $\kappa_{xy}^{\rm sp} = \Delta f_{1}(\Delta/T)/(4\pi)$] by expressing the gradient of the temperature in terms of the gradient of the entropy by using their local-equilibrium relation $\delta s = Tf_{3}(\Delta/T) \delta T/(2\pi v_{\rm F}^2)$. For completeness, ${\bm p}  = T^3 {\bm u} f_{3}(\Delta/T) A(\Delta/T)/(4 \pi v_{\rm F}^4)$.

Eq.~(\ref{eq:bound_cond}) defines the localization length $\ell = \ell(q,\omega)$, which interestingly appears to be a topological property of the system stemming from the presence of the band gap and finite Berry curvature. Importantly, the equation $j_{q,x}(x=0) = 0$ has a solution only if $q \Delta <0$. Recalling that $\Delta$ (which controls the sign of the Berry curvature) has opposite signs in different valleys, we find that only modes with $q<0$ ($q>0$) can propagate in valley ${\bm K}$ (${\bm K}'$). Solving Eq.~(\ref{eq:bound_cond}) we find $\ell^{-1} = -\omega q B(\Delta/T)/\big[T A(\Delta/T)\big]$. Inserting this back into the first line of Eq.~(\ref{eq:NS_equation_0}) and solving for $\omega$ we find
\begin{eqnarray}
\omega &=& |q| \frac{{\tilde v}_{\rm F}}{\sqrt{2}} \left[ 1 + \frac{v_{\rm F}^2q^2}{T^2} \frac{B^2(\Delta/T)}{A(\Delta/T)} \right]^{-1/2}
~,
\end{eqnarray}
which hold for states with $q\Delta < 0$. States with $q\Delta > 0$ cannot propagate. We can at this point reintroduce the viscosity. At small $q$, the boundary conditions and expression for the localization length $\ell$ are unaffected by $\nu$. However, the kinemaric viscosity introduces dissipation of the second-sound modes, and the frequency is replaced in the limit $q\to 0$ by $\omega \to {\tilde v}_{\rm F} |q|/\sqrt{2} + i \nu q^2$. 

It is useful to analyze the behavior of the second-sound dispersion and localization length in the limit of small momentum and temperature. We find that $\omega \to v_{\rm F}|q|\sqrt{T/\Delta}$ and $\ell \to 2 \sqrt{T \Delta}/(v_{\rm F} q^2)$. As Fig.~(\ref{fig:two}b) shows, the localization length decreases at low temperatures, {\it i.e.} the second-sound mode becomes more and more localized at the edge. At the same time, its velocity also decreases, and in the limit of zero temperature it becomes a completely-localized non-dispersive edge mode.

%%%%%%%%%%%%%%%%%%%
\begin{figure}[t]
\begin{center}
\begin{tabularx}{\columnwidth}{X}
\noindent\parbox[c]{\hsize}{\includegraphics[width=0.99\columnwidth]{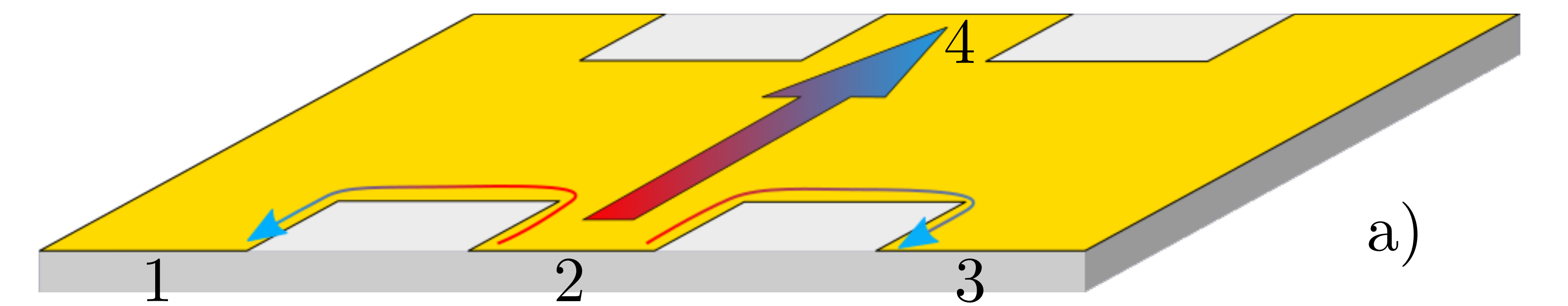}}
%\\
\noindent\parbox[c]{\hsize}{\includegraphics[width=0.99\columnwidth]{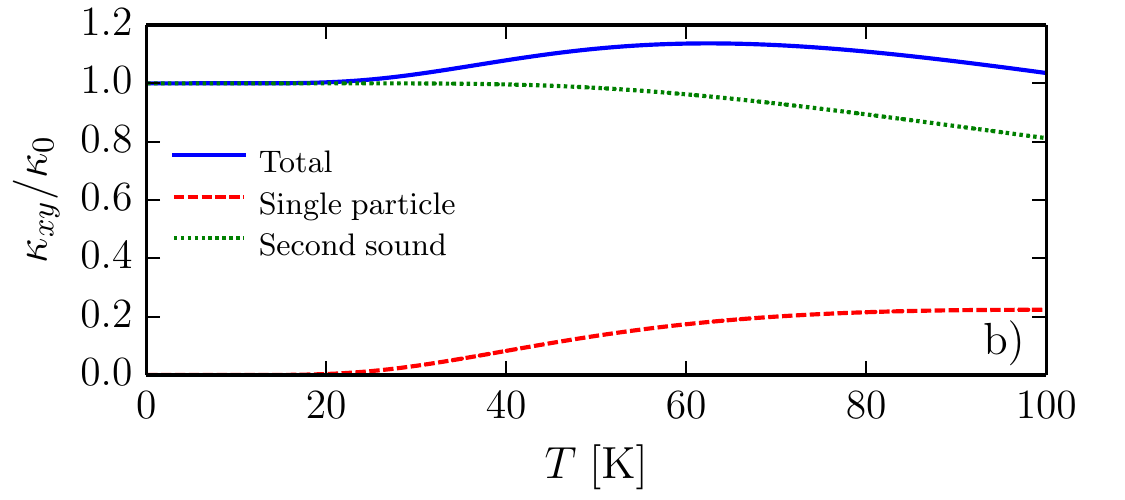}}
\end{tabularx}
\end{center}
\caption{
Panel (a) the proposed experimental setup. A thermal gradient is applied between the middle contacts ($2$ and $4$), and the temperature drop is measured on contacts 1 and 3. The non-local temperature drop can be related to the transverse thermal conductance as in~\cite{Molenkamp_prl_1992,Granger_prl_2009}.
Panel (b) The thermal conductance $\kappa_{xy}$ [in units of the thermal conductance of a perfect one-dimensional bosonic channel $\kappa_0 = \pi^2 k_{\rm B}^2 T/(3 h)$] as a function of the temperature (in ${\rm K}$). We also plot the two components $\kappa_{xy}^{\rm sp}$ and $\kappa_{xy}^{\rm ss}$ in the same units. Note that at low temperature the single-particle contribution is completely suppressed, and the transport is dominated by edge collective modes. At higher temperature the fast growth of $\kappa_{xy}^{\rm sp}$ leads to a change in the slope of the total thermal conductance. 
\label{fig:three}}
\end{figure}
%%%%%%%%%%%%%%%%%%%

{\it Summary and conclusions}---In this paper we studied the problem of thermal transport at the edge of a graphene sheet with broken inversion symmetry. The electron liquid is assumed to be in the hydrodynamic regime~\cite{Landau_Lifshitz_6,Giuliani_Vignale,Torre_prb_2015,Principi_prb_2016,Bandurin_science_2016,Crossno_science_2016}. We find that two counter-propagating, valley-polarized second-sound modes exist when the Fermi energy is in the gap. This result is made possible by the Berry curvature of the band structure~\cite{Xiao_rmp_2010}. Indeed, even though the latter has no impact on the bulk collective modes, it crucially enters into the boundary conditions ensuring that no particle or heat current flows through the edge. Each of the two inequivalent valleys of the Brillouin zone contributes one second-sound collective mode. They propagate in opposite directions along the edge, due to the opposite sign of the Berry curvature.

This result can be experimentally tested in non-local thermal measurements~\cite{Gorbachev_science_2014}. Consider for example the multi-terminal Hall-bar geometry of Fig.~(\ref{fig:three}a). When a thermal gradient is applied, two second-sound modes transport the heat along the edge and the temperature distribution becomes therefore strongly anisotropic. Using known techniques~\cite{Molenkamp_prl_1992,Granger_prl_2009} it is possible to measure the local temperature along the edge, and to relate it to the thermal conductance of the channel. For a {\it bosonic} edge mode, assuming the transmission probability to be exactly one, it is possible to determine the thermal conductance from the Landauer-Buttiker formula,~\cite{Datta_book} {\it i.e.}
\begin{equation}
\kappa_{xy}^{({\rm ss})} = \frac{1}{h} \int_0^{2\Delta} \frac{\omega^2 d\omega }{4 k_{\rm B} T^2 \sinh^2\big[\omega/(2k_{\rm B}T)\big]}
\to \frac{\pi^2 k_{\rm B}^2}{3 h} T
~.
\end{equation}
The final result (the quantum of thermal conductance of a perfect 1D channel) holds in the limit of $k_{\rm B} T \ll 2\Delta$. 

In general, a nontrivial Berry curvature induces also a transverse {\it single-particle} heat current (the so-called thermal Hall effect~\cite{Qin_prl_2011}). The total nonlocal thermal conductance is therefore $\kappa_{xy} = \kappa_{xy}^{({\rm ss})}+\kappa_{xy}^{({\rm sp})}$. However, in the limit $k_{\rm B} T\ll 2 \Delta$, $\kappa_{xy}^{({\rm sp})}$ is {\it exponentially} suppressed~\cite{Qin_prl_2011}. 
The change in the behavior of $\kappa_{xy}$ versus temperature shown in Fig.~(\ref{fig:three}b) is the hallmark of thermal transport mediated by the edge second-sound modes. Another possibility to reveal the signatures of the localized edge modes is via shot noise experiments where one expects that thermal modes will significantly change the Fano factor. This approach was beautifully explored in the context of counter-propagating neutral modes of the fractional quantum Hall effect~\cite{Heiblum_Nat_2010}. 

Apart from graphene, hBN-encapsulated transition metal dichalcogenides, such as WTe$_2$ and MoTe$_2$, that are expected to host time-reversal invariant quantum spin Hall states at monolayer thickness, represent alternative platforms for the observation of the topological collective transport effect we predict in this work.

{\it Acknowledgements}---A.P. and M.I.K. acknowledge support from the ERC Advanced Grant 338957 FEMTO/NANO and from the NWO via the Spinoza Prize. This work was financially supported in part by NSF Grants No. DMR-1606517 and ECCS-1560732 (A.L.). Support for this research at the University of Wisconsin-Madison was provided by the Office of the Vice Chancellor for Research and Graduate Education with funding from the Wisconsin Alumni Research Foundation.

\bibliography{biblio}

\end{document}